\documentclass[traditabstract,letter]{aa}  
\usepackage{graphicx}
\usepackage{txfonts}
\usepackage{natbib}
\bibpunct{(}{)}{;}{a}{}{,} 

\usepackage{color}
	
\newcommand{\Kkms}{%
	\ensuremath{\mathrm{K\,km\,s}^{-1}}}
\newcommand{\kms}{%
	\ensuremath{\mathrm{km\,s}^{-1}}}
\newcommand{\figref}[1]{%
	Fig.~\ref{#1}}

\newcommand{\HII}{%
	H\,{\sc ii}}
\newcommand{\CII}{%
	[C\,{\sc ii}]}
\newcommand{\CI}{%
	[C\,{\sc i}]}
\newcommand\micron{\mbox{$\mu$m}}

\begin{document}
\title{The structure of hot gas in Cepheus B}


\author{B.~Mookerjea\inst{1}
	\and
	V.~Ossenkopf\inst{2}
	\and
	O.~Ricken\inst{2}
	\and
	R.~G\"usten\inst{3}
	\and
	U.~U.~Graf\inst{2}
	\and
	K.~Jacobs\inst{2}
	\and
	C.~Kramer\inst{4}
	\and
	R.~Simon\inst{2}
	\and
	J.~Stutzki\inst{2}
	} 

\institute{Department of Astronomy and Astrophysics, Tata Institute of 
	Fundamental Research, Homi Bhabha Road, Colaba, 
	Mumbai, 400005, India\\
	\email{bhaswati@tifr.res.in}
	\and
	KOSMA, I. Physikalisches Institut, Universit{\"a}t zu
	K{\"o}ln, Z\"ulpicher Str. 77, 50937 K\"oln, Germany
	\and
	Max-Planck-Institut f{\"u}r Radioastronomie, Auf dem H{\"u}gel 69, 53121 Bonn, Germany
	\and
	Instituto de Radio Astronom\'ia Milim\'etrica (IRAM), 
	Avenida Divina Pastora 7, Local 20, 18012 Granada, Spain
} 

\date{Received; accepted}

 
\abstract{
By observing radiation-affected gas in the Cepheus B molecular cloud
we probe whether the sequential star formation in this source is
triggered by the radiation from newly formed stars.
We used the dual band receiver GREAT onboard SOFIA to map \CII{} and
CO 13--12 and 11--10 in Cep B and compared the spatial distribution and
the spectral profiles with complementary ground-based data of low-$J$
transitions of CO isotopes, atomic carbon, and the radio continuum.
The interaction of the radiation from the neighboring OB association
creates a large photon-dominated region (PDR) at the surface of the
molecular cloud traced through the photoevaporation of C$^+$. Bright
internal PDRs of hot gas are created around the embedded young stars,
where we detect evidence of the compression of material and local
velocity changes; however, on the global scale we find no indications
that the dense molecular material is dynamically affected.
}


\maketitle

%

\section{Introduction}

The molecular cloud Cepheus B, located at a distance of about 730 pc can
be regarded as a prototype of a region with sequential star formation.
It has a wedge-like shape, pointing NW to the young
Cepheus OB3 association, including the luminous stars HD 217086 (O7) and
HD 217061 (B1).  The OB association is composed of two subgroups of
different ages, where the youngest group lies closer to the molecular
cloud \citep{Sargent_ApJ_1977}.  The radiation from the association
creates a bright \HII{} region, S155, at the cloud surface. Deeper in
the cloud, star formation continues \citep{Moreno-Corral_A&A_1993},
creating a hot core \citep{Felli_A&A_1978} that was resolved into two
ultracompact \HII\ regions and one extended \HII\ region by
\citet{Testi_A&A_1995}. The main energy source in the hot core is a B1
star in source A of \citet{Testi_A&A_1995}, but more pre-main-sequence
stars are contained in the hot core \citep{Moreno-Corral_A&A_1993}. The
density of molecular material increases further towards the SE,
where two neighboring column density peaks have been identified through
observations of low-$J$ $^{13}$CO and C$^{18}$O
\citep{Ungerechts_ASPConfSer_2000, Mookerjea_A&A_2006}. The molecular
cloud observations by \citet{Beuther_A&A_2000} showed a clumpy structure
that allows a deep penetration of UV radiation from the OB
association and a large-scale velocity gradient with blueshifted
velocities towards the OB association. At a resolution of
43\arcsec, three \CII\ spectra at 158~\micron\ were observed in
Cepheus~B using the KAO \citep{Boreiko_ApJ_1990}. The \CII\ spectra was
found to have two velocity components, one associated with the molecular
material and one with S155. The CO 4-3 and \CI{} observations of
\citet{Mookerjea_A&A_2006} showed that the main heating of the molecular
material occurs in PDRs (photon-dominated regions) created below S~155
and around the hot-spot \HII\ regions. The FUV field is estimated to be
between 1500 to 25$\chi_0$, where $\chi_0$ is the average interstellar
radiation field \citep{Draine_ApJS_1978}.

To look for indications of sequential star formation due to radiative
triggering as proposed by \citet{Getman2009}, the density, temperature
structure, and kinematics of the FUV irradiated PDRs at the cloud
surfaces need to be probed.  Signs of triggering processes would be the
formation of density enhancements correlated with the radiative pressure
increase at the cloud surface \citep[radiative shells, see
e.g.][]{Dale_MNRAS_2009} or indications of velocity gradients in the
molecular material created below the radiative compression zones.

To address these questions, we performed spectral mapping observations
of Cep B in \CII{} and CO 13-12 and 11-10 using the German REceiver for
Astronomy at Terahertz Frequencies
\citep[GREAT][]{GREAT}\footnote{GREAT is a development by the MPI
f\"ur Radioastronomie and KOSMA/Universit\"at zu K\"oln, in cooperation
with the MPI f\"ur Sonnensystemforschung and the DLR Institut f\"ur
Planetenforschung.} onboard the Stratospheric Observatory For Infrared
Astronomy (SOFIA). 
The new data were combined with existing observations of low-$J$ CO
isotopologs and \CI\ to  probe the structure of the PDR and the
velocity distribution.


\section{Observations}

We observed the hot core region of Cepheus B during the nights of April
7 and July 19, 2011 with GREAT on SOFIA at an altitude of $\sim
42000$ feet. During the two nights were observed areas of the same size
(2\farcm4$\times$1\farcm6), but offset relative to each other.
 GREAT is a modular heterodyne instrument, with two
channels used simultaneously. In April, channel {\bf L1\#a} was tuned to
CO 11-10 (1267.015\,GHz) and channel L2 was tuned to \CII\
$^2$P$_{3/2}\rightarrow ^2$P$_{1/2}$ (1900.537\,GHz) for four coverages
of the region. In July, {\bf L1\#b} was tuned to CO 13-12 (1496.923\,GHz),
while L2 was tuned to \CII\ for three coverages. The HPBWs are 21\farcs3
at 1.3\,THz, 19\farcs6 at 1.5\,THz, and 15\arcsec\ at 1.9\,THz.

The observations were performed in position-switched on-the-fly
(OTF) mode with an OFF position at $\Delta \alpha = -600\arcsec ; \Delta
\delta =+600\arcsec $.  As backend we used the acousto optic
spectrometer with a bandwidth of 1\,GHz and a resolution of 1.6\,MHz.
The integration time was 0.6 minutes/position. The measured system
temperatures were 3100\,K for CO 11-10, 2500\,K for CO 13-12, and
3200-3500\,K for \CII. The forward efficiency is $\eta_f=0.95$ and the
main beam efficiencies ($\eta_{\rm mb}$) are 0.51 for \CII\ and 0.53 for
CO 11-10 lines \citep{Guan2012}.  Details of calibration of the data are
presented by \citet{Guan2012}. We fitted and subtracted baselines
of the fourth or sixth order to the \CII\ and CO data.

Complementary observations of \CI\ ($^3$P$_1\rightarrow^3$P$_0$ at
492.161\,GHz)  were performed in December 2002 using the CHAMP
2$\times$8 pixel array receiver \citep{Guesten_SPIE_1998} at the CSO.
The fully sampled \CI\ map of 3\farcm3$\times$3\farcm7 was
observed in position-switched mode (reference position -300\arcsec, 0).
The HPBW for \CI\ was 14\farcs5 at 492 GHz and $\eta_{\rm MB}$ was
0.51 \citep{Philipp_A&A_2006}.  The \CI\ data were smoothed to
20\arcsec\ for comparison.


\section{Morphology}

\begin{figure}[!h]
\centering
\includegraphics[width=0.95\linewidth]{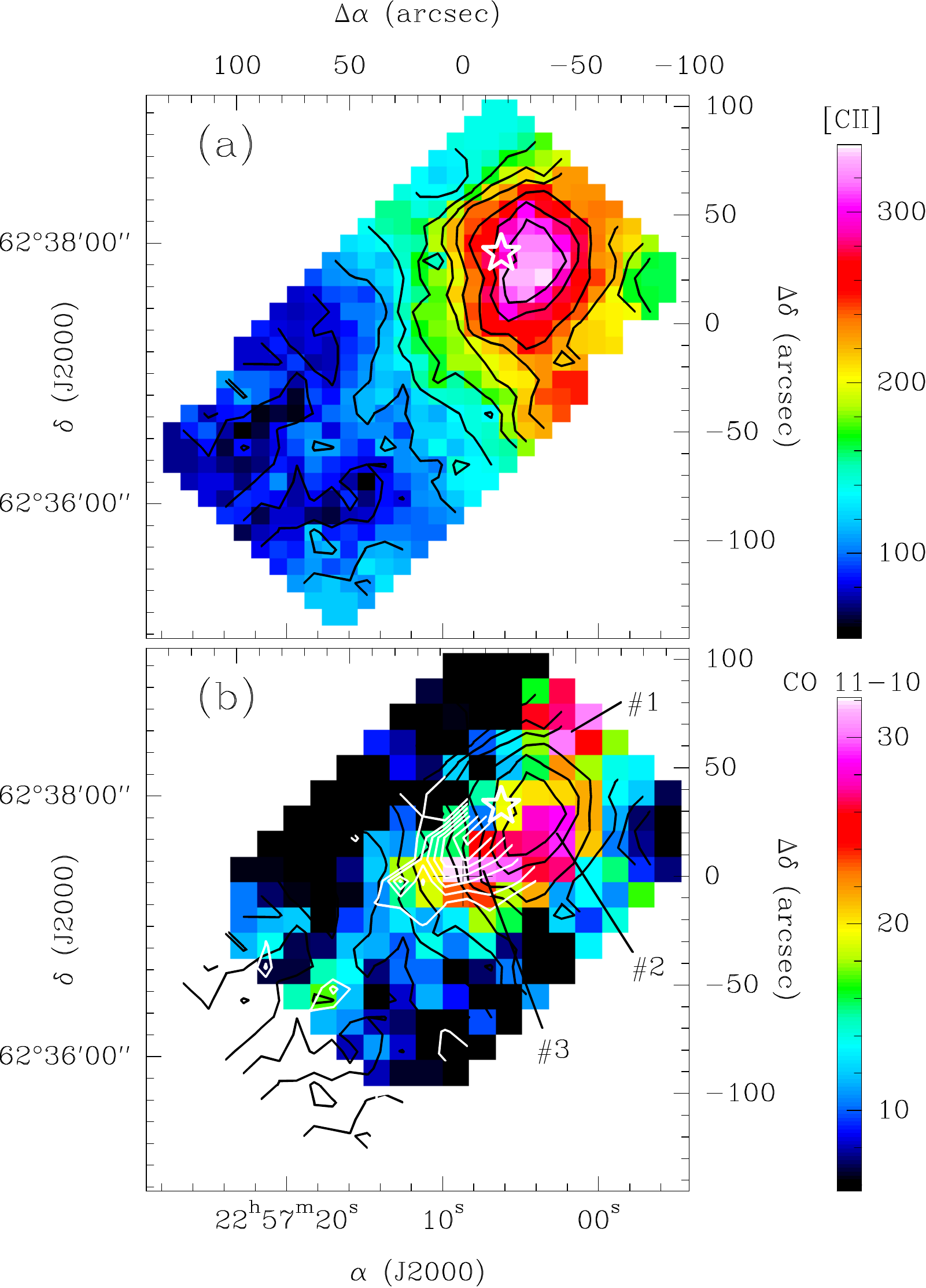}\vspace*{-0.33cm}
\caption{{\bf (a)} Velocity-integrated intensity map of \CII\ in 
color and contors. The contor levels are at 79 to 339.4 ~\Kkms\
 (peak) in steps of 29~\Kkms.  {\bf (b)} Integrated intensity map of CO 11-10
(color) overlayed with contors of CO 13-12
({\em white}) and \CII\ ({\em black}) intensities. Contour levels for CO
13-12 are: 2.1 to 11.7~\Kkms\ (peak). $ v_{\rm LSR} $-range of integration is 
\CII{}: $ \left[ -20;-5 \right] $, CO: $ \left[ -16;-11 \right] $ \kms .
The white star denotes the B1 star at
$ ( \alpha ; \delta ) = ( 22^h57^m6.2^s; 62\degr 37\arcmin 55.4\arcsec ) $. 
Offsets of the marked CO peaks \#1, \#2, and \#3 are
(-45\arcsec,+64\arcsec), (-41\arcsec,+20\arcsec), and
($-9$\arcsec,+3\arcsec), respectively.
\label{fig:integrated_intensities}} 
\end{figure}

Figure~\ref{fig:integrated_intensities} shows the integrated intensity maps
of \CII\ and \mbox{CO 11-10} and \mbox{CO 13-12}. The (0,0) position of
the maps is at $ \alpha = 22^h57^m8$\fs$7 $ and $ \delta = 62\degr
34\arcmin 23\arcsec $ (J2000). The \CII\ emission shows a single peak
(-34\farcs7,-22\farcs9)
located to the W of an embedded B1 star, and the emission drops 
towards the E. The CO 11-10 exhibits three peaks
referred to as \#1, \#2 and \#3 in
the rest of the paper, forming a shell around the \CII{} peak. The
emission is extended further in the NW-SE direction.

\begin{figure}[!h]
	\centering
	\begin{minipage}[b]{.49\linewidth}
		\raggedright \includegraphics[width=.98\linewidth]{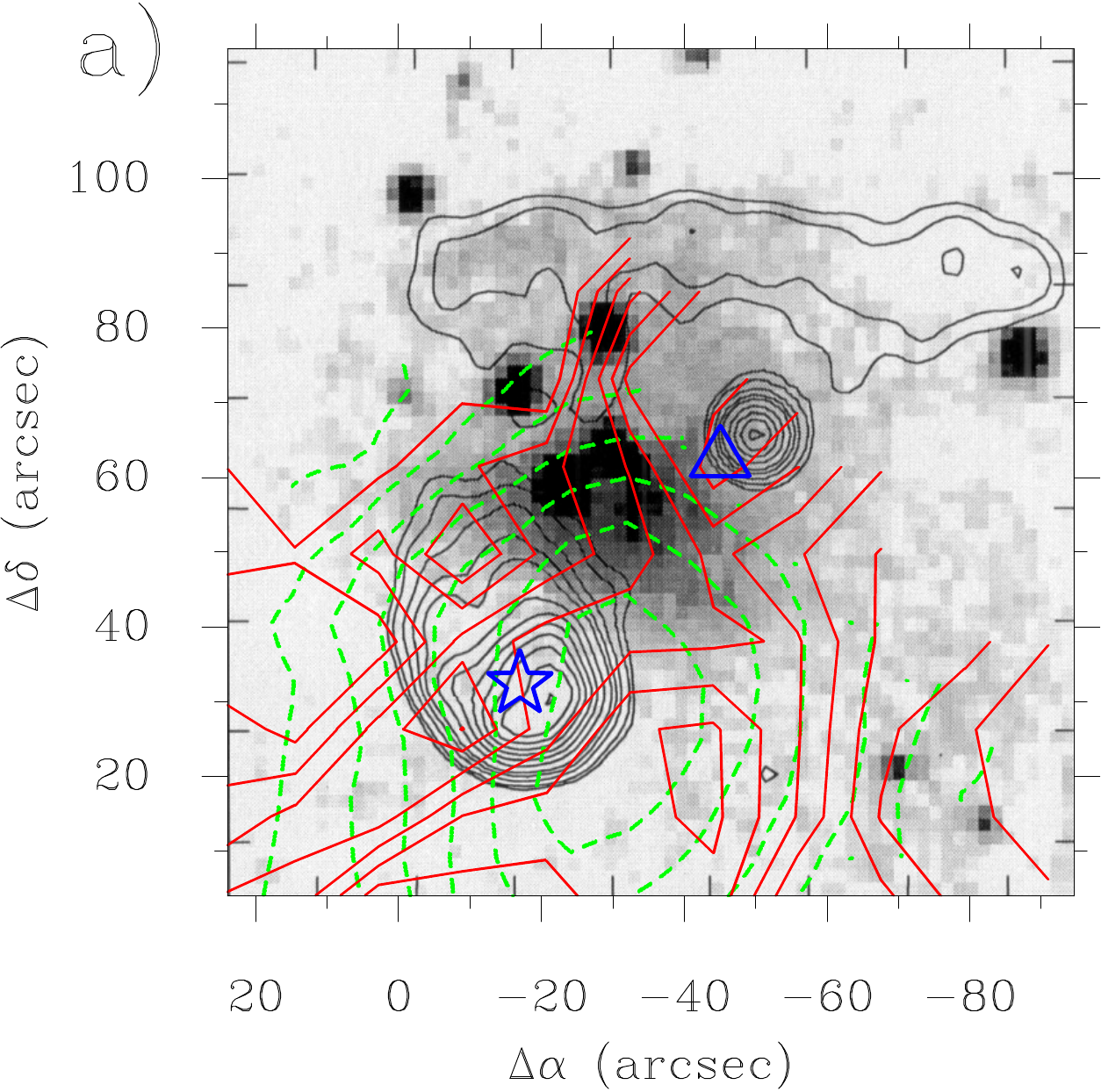}
	\end{minipage}
	\begin{minipage}[b]{.49\linewidth}
		\raggedleft \includegraphics[width=.98\linewidth]{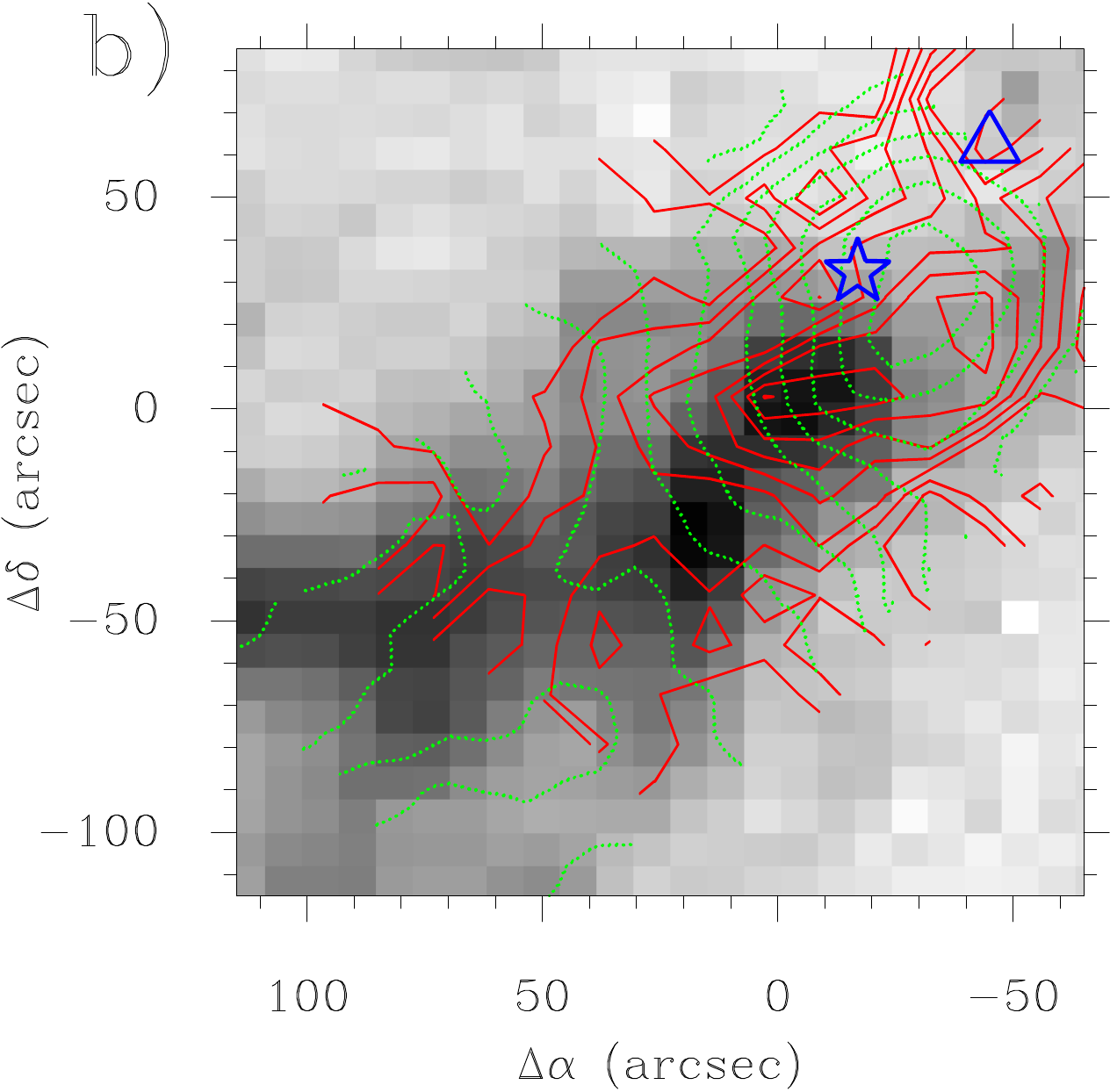}
	\end{minipage}
	\caption{{\bf (a)} Optical image (\emph{grayscale}) overlayed with 
	contors of 8.4~GHz continuum emission (\emph{black}) 
	\citep{Testi_A&A_1995} and 
	contors of \CII\ (green dashed) and CO 11-10 (red). {\bf (b)}~
	Intensity of $ ^{13} $CO 2-1 emission integrated between $v_{\rm
	LSR} = -17$ to $-10$~\kms\
	\citep{Ungerechts_ASPConfSer_2000} (\emph{grayscale}) overlayed with
	contors of CO 11-10 (\emph{red}) and \CII\ ({\em green dotted}). 
	Contour levels of \CII\ are the same as in 
	\figref{fig:integrated_intensities}, for CO 11-10: 6.8 to
	32.1~\Kkms\ (peak) in steps of 3.6 \Kkms. The star and the
	triangle denote the positions of Testi's NIR sources A and B, respectively. 
	\label{fig:radio_13CO}}
\end{figure}

Figure~\ref{fig:radio_13CO}a compares the \CII\ and CO 11-10
observations with radio continuum observations by
\citet{Testi_A&A_1995}. Testi et al. show that the 8.4~GHz radio
continuum emission component A, a blister \HII\ region, peaks slightly
to the SW of the position of the B1 star. We find a sequence of
peaks in that direction starting with the radio continuum close to the
heating star, the \CII\ peak, and the CO 11-10 peak \#2, consistent with
the expected layered structure of a PDR.  The CO 11-10 peaks \#2 and
\#3, both S of the source A trace dense and warm molecular gas. Peak
\#1 coincides with the radio source B \citep{Testi_A&A_1995}. The CO
13-12 emission appears to be centered close to the B1 star, but the
detailed distribution around it could not be observed due to time
constraints. 

Figure~\ref{fig:radio_13CO}b compares our data with IRAM 30m $^{13} $CO 2-1
observations at 10$''$ resolution by \citet{Ungerechts_ASPConfSer_2000},
tracing molecular gas with high column densities, which has a ridge like
structure extended towards the main cloud E of the map. A similar
intensity distribution is also seen in the $^{13} $CO 1-0 map
\citep{Ungerechts_ASPConfSer_2000}, the \CI\ 1--0 map (see
Fig.~\ref{fig:PDR_strip}), and in the CO 4-3 map \citep[at
1\arcmin][]{Mookerjea_A&A_2006}.  The CO 11-10 peak \#3 coincides with a
column density peak in these maps, while peaks \#1 and \#2 indicate
an increased CO excitation, e.g. due to a higher temperature of the
molecular material.
More PDR layers are formed at the surface of the molecular material
in the SE. They are also traced by the \CII{} and the high-$J$
CO lines but at a low level of emission.  We detect almost no CO
emission to the N and NE of the embedded B1 star.

Since \CII\ can also originate in ionized gas, it is reasonable to
expect enhanced \CII\ emission at the location of the radio continuum
peaks detected by \citet{Testi_A&A_1995}.  However, the \CII\ emission
shows no enhancement at the location of the CO peak \#1, which coincides
with the B-component.  This indicates that the source creating the
B-component of the radio continuum emission ionizes at most a small
fraction of the molecular material, consistent with Testi's suggestion
that source B is a protostar, not yet ionizing, but heating up its
surroundings.  

Our observations are consistent with Testi's suggestion of a
blister-type \HII -region for component A, created by the B1 star, and a
density gradient of the surrounding material leading to an ``open cone''
configuration.  The density decreases to the N, so the cone opens in
this direction.  The ionized material may escape to the N, but cannot
expand much to the S because of denser molecular material there.  This
results in increased and sharply bounded radio continuum emission to the
S and in extended, but lower intensity emission to the N of the position
of the embedded star.


\section{PDR layering}
\begin{figure}[!h]
	\centering
	\includegraphics[angle=270,width=8.3cm]{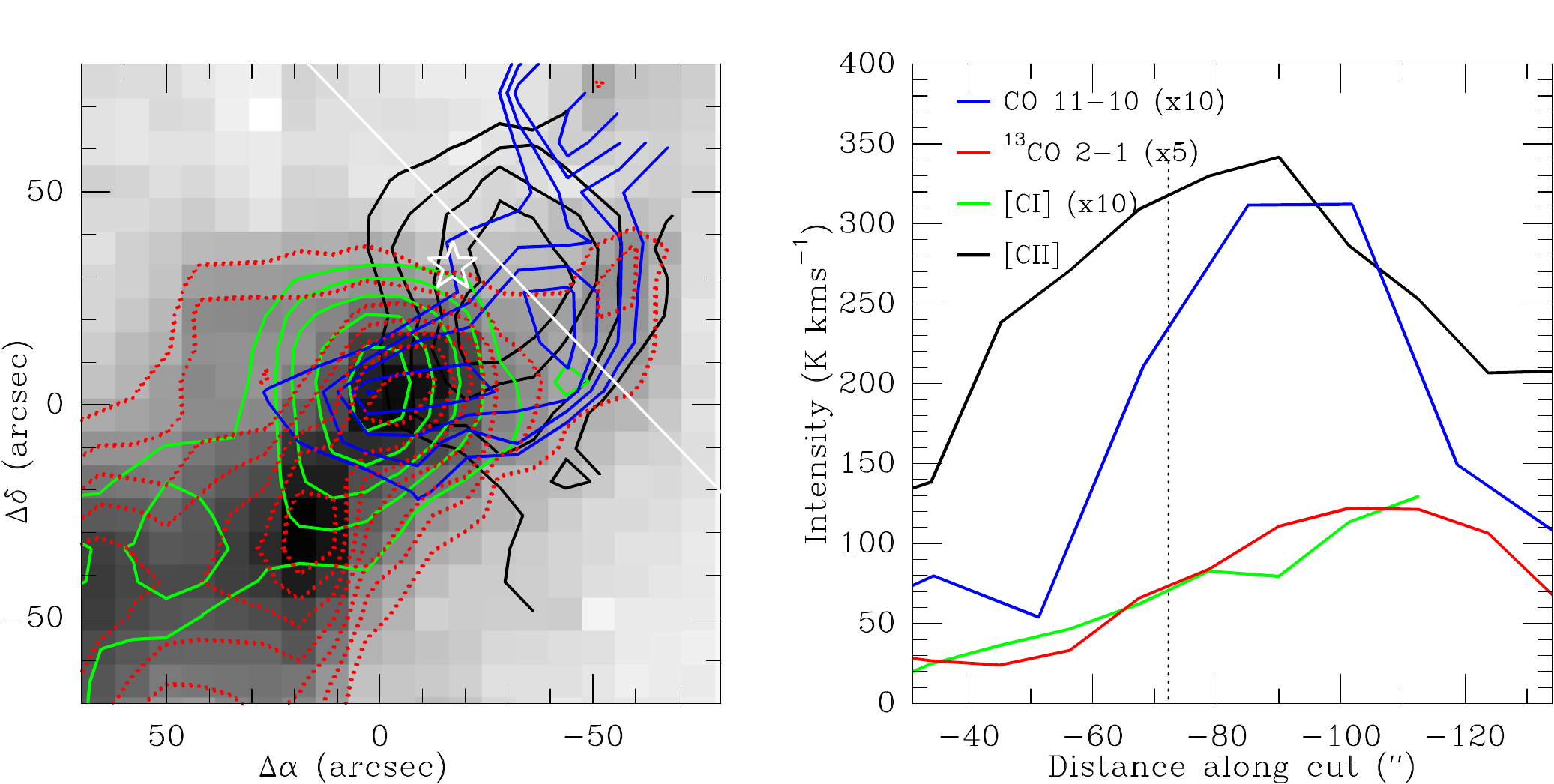}
	\caption{{\it Left:} Map of integrated intensity distributions of 
	$ ^{13} $CO 2-1 (greyscale, red contor), \CII{} (black), CO
	11--10 (blue),
	and \CI{} (green). {\it Right:} Cut through the \CII{} peak and
	the CO 11--10 peak \#2. The black dotted vertical line denotes
	the position closest to the B1 star along the cut. Integration
	ranges: \CII\ and \CI\ $ \left[ -20;-5 \right] $, CO 11--10, and $ ^{13} $CO $ \left[ -17;-10 \right] $ (\kms ). }
	\label{fig:PDR_strip}
\end{figure}

As discussed above we find a layering sequence of \CII, high-$J$ CO and
low-$J$ CO peaks around the \HII\ region associated with the B1 star.
Figure~\ref{fig:PDR_strip} shows the effect of this layering in terms of
the mutual shift of the peaks of the \CII{}, CO 11--10, and $^{13}$CO
2-1 emission.  We find a separation of about $0.02$~pc between \CII{}
and CO 11--10 and of $0.08$~pc between \CII{} and $^{13}$CO 2-1. For
\CI{} the peak emission is seen at the edge of the mapped region. The
\CI\ distribution is similar to that of $^{13}$CO 2-1, but the emission
is weak in the considered cut.  The shift in the peaks is consistent
with the predictions of a uniform PDR model, although in reality, it
will also be affected by small-scale density variations.

For a qualitative comparison we used the KOSMA-$ \tau $ PDR model
\citep{Roellig_A&A_2006} to simulate gas temperature and chemical
structure inside the PDR.  KOSMA-$ \tau $ considers a spherical clump
illuminated by an isotropic FUV radiation field and cosmic rays. Input
parameters are the radiation field, the clump mass and the density at the
clump surface.  It assumes that the clump has a power-law density profile of $
n(r) \propto r^{-1.5} $ for $ 0.2 \leq r/r_{cl} \leq 1 $ and $ n(r) =
\mathrm{const.} $ for $ r/r_{cl} \leq 0.2 $. To simulate a
plane-parallel PDR through a spherical model, we chose a clump 
of $ 10^3\,\mathrm{M}_{\sun}$, i.e. with a mass much larger than the 
available molecular mass \citep[see][]{Mookerjea_A&A_2006}.  The outer 
layers of this massive clump  then form an almost plane-parallel configuration
suitable to studying the observed stratification.

\begin{figure}[!h]
\begin{center}
	\includegraphics[width=6.7cm]{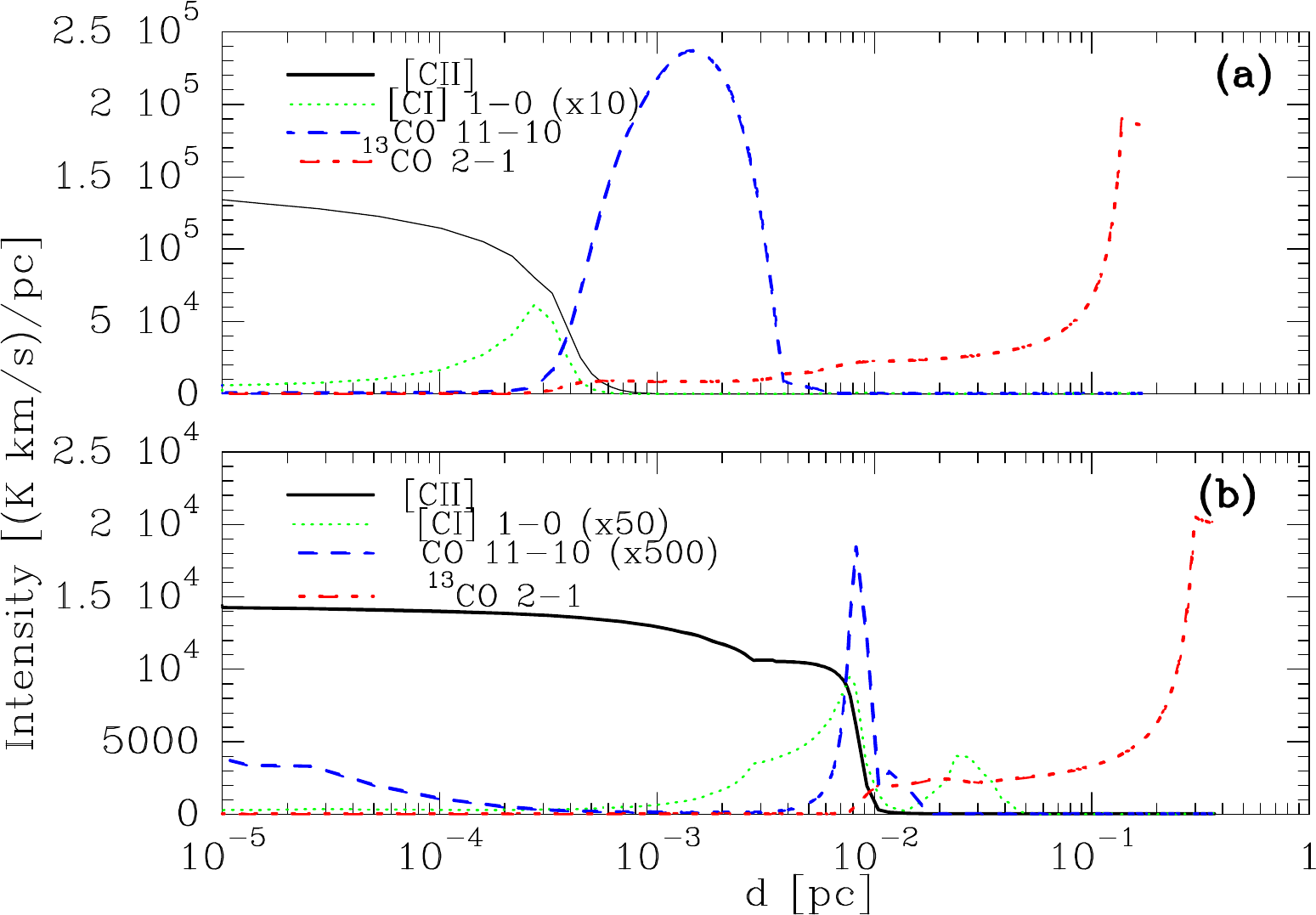}
	\caption{Optically thin line intensities per cloud depth for the two
	model densities: {\bf (a)} $ n_{\rm surf} = 10^6\,\mathrm{cm}^{-3}
	$, {\bf (b)} $ n_{\rm surf} = 10^5\,\mathrm{cm}^{-3} $.
	Radiation field and clump mass are $ 1000\,\chi_0 $ and $
	1000\,\mathrm{M}_{\sun}$, respectively.}
	\label{fig:KOSMA-tau_int}
\end{center}	
\end{figure}

Figure~\ref{fig:KOSMA-tau_int} shows the local optically thin emissivity,
i.e.  the intensities per cloud depth without optical depth correction,
for two models with a density $ n(r_{cl} =
10^5\,\mathrm{cm}^{-3} $ and $ n(r_{cl} = 10^6\,\mathrm{cm}^{-3} $
illuminated by a radiation field of $ 1000\,\chi_0 $
\citep{Mookerjea_A&A_2006}. The model predicts a stratification with a
0.001\dots 0.01~pc separation of the CO 11--10 peak relative to the
\CII{} peak and a further separation of 0.12 \dots 0.3~pc to the
$^{13}$CO 2--1 peak\footnote{$^{13}$CO 2--1 always shows a
broad peak filling a large fraction of the clump core, independent
of the clump mass, indicating that the displacement of the $^{13}$CO 2--1
peak relative to the PDR surface is fairly well determined by the density
structure than by illumination.}, indicating that the parameters used are reasonable,
but do not provide a full match to the observations. The observed
intensity ratios of the three lines suggest that the density in the
cloud is closer to $10^6\,\mathrm{cm}^{-3}$ than to
$10^5\,\mathrm{cm}^{-3}$.  Since the CO 11-10 line is very sensitive to the
gas density, the similar peak intensities at \#2 and \#3 indicate
approximately equal gas densities, while the higher $^{13}$CO and \CI{}
intensities at peak \#3 indicate higher column densities of molecular
gas there. 

The model, however, predicts a narrow, but weak \CI{} peak in front of
the CO 11--10 layer, while the \CI{} emission is very extended on the map
\footnote{Spot checks with the PDR models by M.  Kaufman and the Meudon
group show the same inversion in the \CI{} layering that we find with
the KOSMA-$\tau$ model.}. In contradiction to the model, the \CI{}
emission seems to measure the molecular column density, so not following the 
stratification of the other tracers.


\section{Velocity structure}

\begin{figure}[!h]
	\centering
	\includegraphics[width=\linewidth]{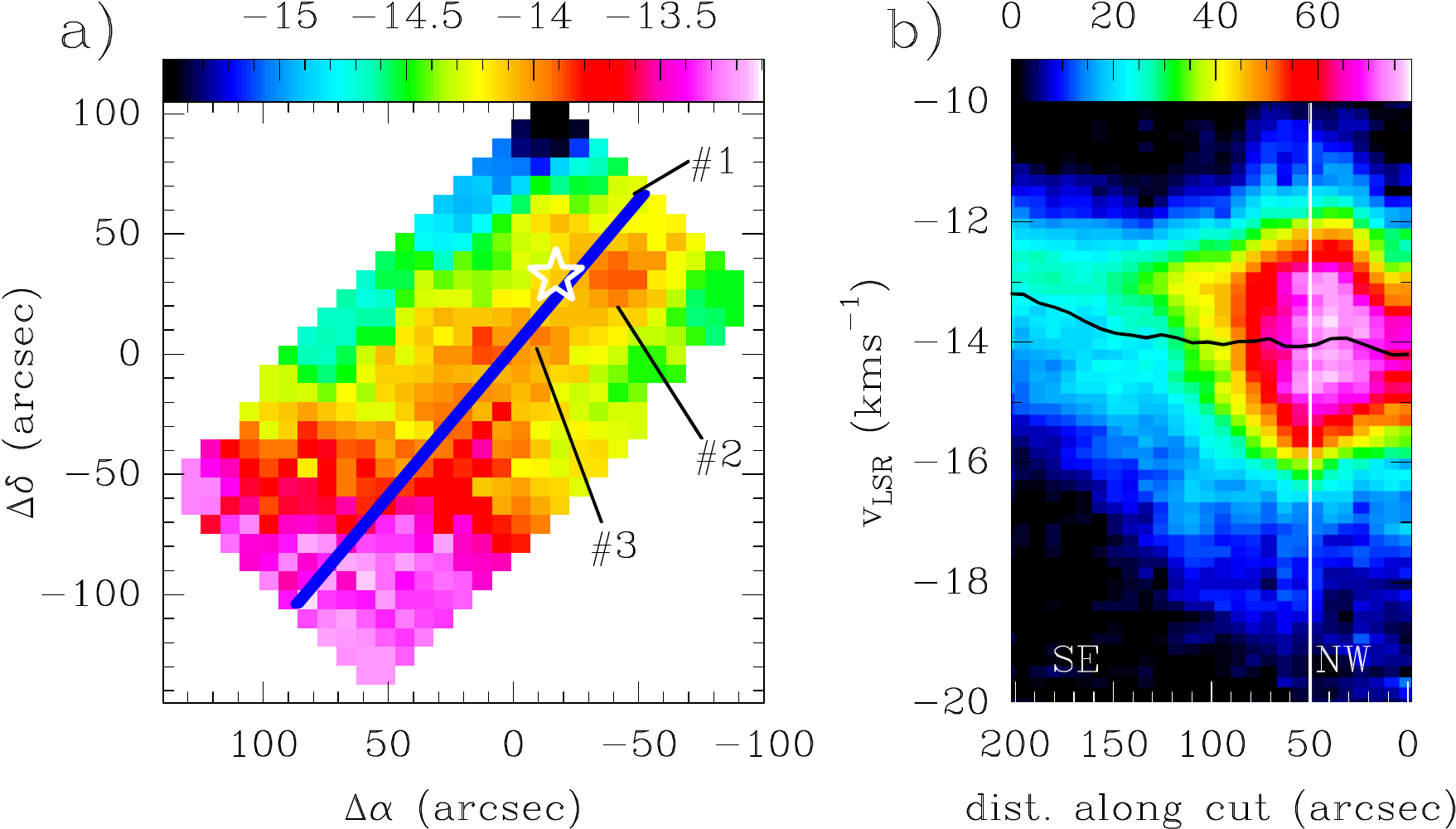}
	\caption{{\bf (a)}~Map of the $ v_{\rm LSR} $ values (in \kms{}) for Gaussian
	fits to the \CII\ spectra. {\bf (b)}~PV diagram for \CII\ along the cut
	indicated in {\bf (a)}. The white line shows the position of the
	\CII\ peak and the solid black line the centroid velocity 
	along the cut.}	
	\label{fig:pvdiag}
\end{figure}

To study the main characteristics of the \CII\ velocity distribution we
show in \figref{fig:pvdiag} the map of the centroid velocities and the
corresponding position-velocity diagram for a cut through the map.  The
map of the centroid velocities was generated by fitting a single-component Gaussian
profile to each spectrum. The velocity distribution is dominated by a
large-scale gradient of about 1~\kms\ over 200\arcsec\ from NW
to SE, visible as a color gradient in \figref{fig:pvdiag}a and as
a global slope in \figref{fig:pvdiag}b. This is consistent with the
velocity gradient of $4.2 \times 10^{-3}$~\kms{}/arcsec in the molecular
cloud derived from the channel maps of CO 3--2 observations
\citep{Beuther_A&A_2000}.  Moreover, one can see a good correlation
between the centroid velocity map and the $^{13}$CO map in
\figref{fig:radio_13CO} in the sense that pixels associated with strong
emission in low- and mid-$J$ CO have systematically higher velocities
than nearby pixels measuring lower density material. The velocity of the
low-density material is closer to that of S155; i.e., the ionized
gas is always somewhat blueshifted relative to the denser molecular
cloud.

The position-velocity diagram \figref{fig:pvdiag}b) shows a broadening
of the \CII\ line at the position of the \CII\ peak and there appears to
be an additional velocity component visible as a blue shoulder.  An
interesting feature is, however, the change in the velocity gradient
around the \CII\ peak. The contors in the PV diagram form a
parallelogram NW of the peak where the large-scale velocity gradient is
inverted. The inversion is not very pronounced when considering the
Gauss-fit velocities along the cut, but is apparent as higher velocity
in the map (\figref{fig:pvdiag}a) around CO peak \#2.

\begin{figure}[!h]
	\centering
	\includegraphics[width=5.7cm]{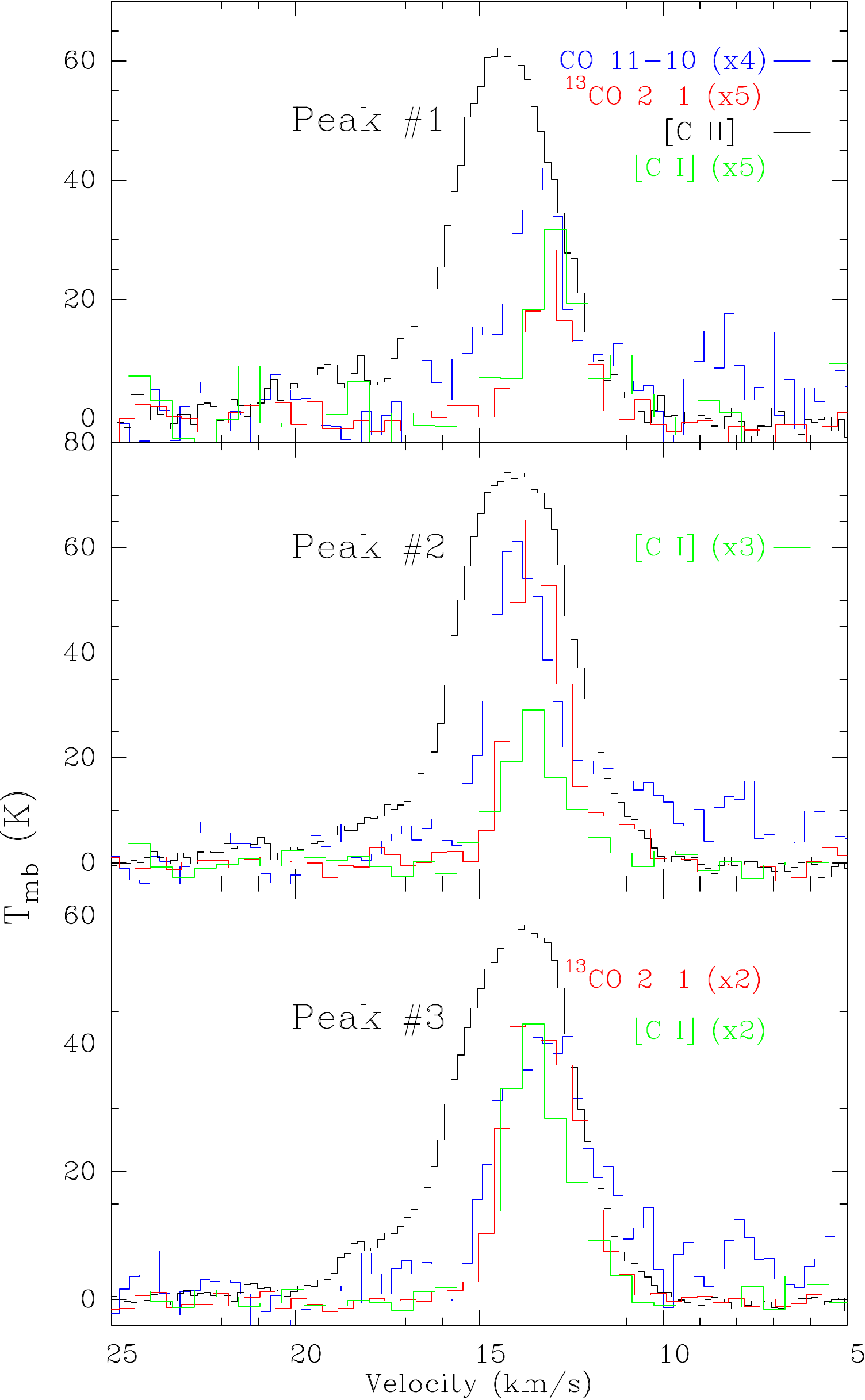}
	\caption{Spectra averaged over an area of 33\arcsec\
	diameter around the CO 11-10 peaks \#1, \#2, and \#3.}
	\label{fig:spectra_max1_max2}
\end{figure}

To look at the details of the spectral profiles, we averaged the
spectra around the peak positions \#1, \#2, and \#3 over an area of
33$''$ diameter (\figref{fig:spectra_max1_max2}).  We find emission from
different velocity components. The normal molecular cloud material at
velocities of $-13.5$\dots $-13$ \kms{} is traced by $^{13}$CO 2-1,
\CI, and CO 11-10 in \#1 and \#3.   In this complementary analysis
approach, involving positions not all on the cut discussed above, we
find that at \#2, CO 11-10 is, significantly blueshifted, showing a
local inversion of the velocity gradient similar to the inversion seen
in \CII{} along the cut. This means that the embedded B1 star 
not only creates a local \HII{} region, but it also affects the
dynamics of the hot gas in the PDR around the \HII{} region on a scale
of about 30$''$, i.e. 0.1~pc. 

The \CII{} line is much wider at all positions. It includes the velocity
of the molecular cloud but also an additional blueshifted component with
velocities of about $-16$~\kms{} and a wing extending down to
$-21$~\kms{}.  \citet{Boreiko_ApJ_1990} showed that the $-16$~\kms\
component matches the $v_{\rm LSR}$ of the H$\alpha$ emission from the
\HII\ region S155.  At \#1, this $-16$\,\kms\ is the dominant component,
showing that most of the \CII{} emission stems from the \HII{} region
there. The broad blue wing is, however, also present at \#2 and \#3,
and actually strongest at \#3. This may represent ionized material
that is ablating from the cloud surface. Depending on the exact geometry
of cloud and main radiation source, we find two  possible
interpretations. If HD\,217086, located slightly behind the cloud
\citep{Felli_A&A_1978}, represents the main sources of UV photons, 
the acceleration of low-density material through the
radiation pressure could lead to the wing of blue shifted material.  If
HD 217061, in front of the cloud, is the main radiation source, we 
could be tracing the signature of a photo-evaporation flow from the surface as
predicted by \citet{GortiHollenbach2002}. Detailed analysis
confirming either of the two possibilities is beyond the scope of this
paper.


\section{Conclusion}

The \CII{} emission is affected on large scales by radiation from the
Cepheus OB3 association with a gradual change from material stemming
mostly from the S155 \HII{} region in the NW through material
photoevaporating from the molecular cloud surface, and since it is
potentially accelerated through radiation pressure, to PDR material
within the molecular cloud that is ionized through the FUV radiation,
but resides at the velocities of the parental cloud with the known
large-scale velocity gradient.  For the tracers of the denser material,
we find no large-scale effects but local velocity shifts due to the
embedded ultracompact \HII{} regions.  They create a partially inverted
velocity gradient, a local broadening of the line profiles and disperses
the material around them, but do not create any long-range effects.

Overall, we found no indications that a radiative triggering mechanism
creates the sequential star formation in Cepheus B. Radiative
star-formation triggering may work on very small scales around the
\HII{} regions mostly unresolved here, but cannot explain a large-scale
sequence.
\begin{acknowledgements}
We thank Sabine Philipp for providing us with the 
CHAMP \CI{} data and Hans Ungerechts for providing the IRAM $^{13}$CO data. 
This paper is based on observations made with the NASA/DLR Stratospheric
Observatory for Infrared Astronomy. SOFIA Science Mission Operations are
conducted jointly by the Universities Space Research Association, Inc., under
NASA contract NAS2-97001, and the Deutsches SOFIA Institut under DLR
contract 50 OK 0901. We thank the SOFIA engineering and operations teams for their
support during Early Science. The research presented here was supported by the German
\emph{Deut\-sche For\-schungs\-ge\-mein\-schaft, DFG\/} through project
number SFB 956C. This research made
use of  the SIMBAD database, operated at the CDS, Strasbourg,
France, and 
NASA's Astrophysics Data System Bibliographic Services.
\end{acknowledgements}

\bibliographystyle{bibtex/aa}

\end{document}